\documentclass[twocolumn,showpacs,amsmath,amssymb,prl,floatfix]{revtex4}

\usepackage{graphicx}
\usepackage{dcolumn}
\usepackage{bm}
\usepackage{bbm}
\usepackage{multirow,amssymb,amsbsy,amsmath}
\usepackage{stmaryrd}

\newcommand{\ddim}{\udelta\kern0.1em}

\newcommand{\beikonst}[2]{\left( #1 \right)_{\kern-0.2em #2}}

\newcommand*{\bra}[1]{\mathopen{\langle}#1\mathclose{|}}
\newcommand*{\ket}[1]{\mathopen{|}#1\mathclose{\rangle}}

\newcommand{\s}{{\sigma}}

\newcommand{\heff}{{H}^{}_{\mathrm{eff}}}

\newcommand{\ketbrap}[2]{\mathopen{|}#1\mathclose{\rangle}\hspace{-0.25em}\mathopen{\langle}#2\mathclose{|}}

\begin{document}

\preprint{APS/123-QED}

%
%

\title{Two-stage melting in systems of strongly interacting Rydberg atoms}

\author{Hendrik\ Weimer}%
\affiliation{Institute of Theoretical Physics III, Universit\"at Stuttgart, %
              70550 Stuttgart, Germany}%
\email{hweimer@itp3.uni-stuttgart.de}%
\author{Hans\ Peter\ B\"uchler}%
\affiliation{Institute of Theoretical Physics III, Universit\"at Stuttgart, %
              70550 Stuttgart, Germany}%

\date{\today}%

\begin{abstract}
We analyze the ground state properties of a one-dimensional cold
atomic system in a lattice, where Rydberg excitations are created by
an external laser drive. In the classical limit, the ground state is
characterized by a complete devil's staircase for the commensurate
solid structures of Rydberg excitations. Using perturbation theory and
a mapping onto an effective low energy Hamiltonian, we find a
transition of these commensurate solids into a floating solid with
algebraic correlations.  For stronger quantum fluctuations the
floating solid eventually melts within a second quantum phase
transition and the ground state becomes paramagnetic.
\end{abstract}


\maketitle

The quantum melting of solids is a paradigm for the manifestation of
quantum fluctuations at zero temperature. This quantum phase transition
is driven by the competition between the interaction energy giving
rise to a crystalline structure with localized particles and the
kinetic energy preferring a delocalization of the particles. The
qualitative behavior of the transition can be described by the
Lindemann criterion, which states that the solid melts if the
fluctuations around the mean position reach a certain fraction of the
lattice spacing \cite{Kleinert1989}. The zero temperature melting of
solids with purely repulsive interactions has been studied in detail
for Wigner crystals \cite{Wigner1934}, and has recently
attracted large attention for dipolar system realized with polar
molecules \cite{Buchler2007} and Rydberg gases
\cite{Weimer2008a,Pohl2010,Schachenmayer2010}. In this Letter we study
the quantum melting of crystalline phases in driven Rydberg systems.

Rapid experimental progress motivated by coherent applications like
quantum computing \cite{Saffman2010} and quantum simulation
\cite{Weimer2010} has pushed the field of Rydberg atoms from
single-particle physics into an area where strong many-body effects
are important
\cite{Heidemann2007,Urban2009,Gaetan2009,Younge2009,Pritchard2009,Schempp2010}.  Within
these driven strongly interacting systems, crystalline many-body
ground states have recently been proposed
\cite{Weimer2008a,Pohl2010,Schachenmayer2010}, where an ordered
arrangement of Rydberg excitations spontaneously appears in a
translationally invariant configuration of frozen atoms. The absence
of a kinetic term in the microscopic Hamiltonian describing these
systems requires a deeper analysis to fully understand the nature of
the quantum melting of these crystalline structures. 
Here, Rydberg atoms in one-dimensional (1D) lattices
\cite{Olmos2009} are particularly suitable to investigate the
properties of Rydberg crystallization.

\begin{figure}[tb]
  \includegraphics[width=0.9\linewidth]{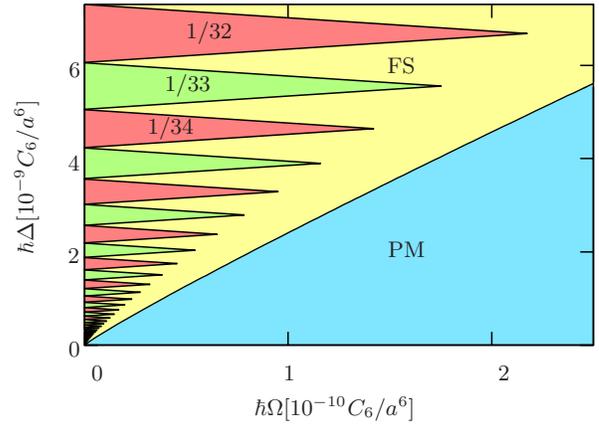}
  \caption{Ground state phase diagram: We find commensurate solids with filling $f=p/q$ 
    describing a complete devil's staircase for $\Omega=0$.  The lobes with $p> 1$ are not visible 
    within this scale.For increasing Rabi frequency $\Omega$, a quantum phase transition takes place into a gapless
    floating solid (FS) with algebraic correlations for the solid structure and in general incommensurate density, 
   and eventually a second melting transition occurs into a gapped paramagnet (PM).
   This second transition line satisfies $\Delta \sim \Omega^{12/13}$.
   }
  \label{fig:phase}
\end{figure}

In this Letter, we develop an effective theory allowing to study the
quantum melting of crystalline phases of strongly interacting and
driven Rydberg atoms in 1D lattices and determine the complete phase
diagram.  In contrast to the conventional melting of crystalline
solids with the quantum fluctuations driven by the kinetic energy,
here, the quantum fluctuations derive from the competition between
interaction energy and the external laser drive, coupling the atomic
ground state to the excited Rydberg state.  Deriving an effective
low energy description, we find a two-stage melting: first from a
commensurate solid with true long-range order to a floating solid with quasi long-range order, and
finally to a paramagnetic phase, see Fig.~\ref{fig:phase}.  The latter
transition from the floating solid to the paramagnet is in
agreement with predictions from a universal scaling function
\cite{Low2009}.

We start with the microscopic Hamiltonian describing the driven
Rydberg system: Rydberg excitations are created by two-photon
processes via an intermediate atomic level that can be adiabatically
eliminated. Consequently, the internal structure of the atoms can be
treated as a single hyperfine ground state $\ket{g}$ and a
single Rydberg level $\ket{e}$, giving rise to a spin $1/2$
description. The coupling between the two states by an external laser
field is given by the Rabi frequency $\Omega$ and the detuning
$\Delta$.  The interactions between the Rydberg states are described by a
repulsive van der Waals interaction with a $C_6$ coefficient.  Here, we focus on a 1D setup, where the atoms are trapped in a
lattice with lattice spacing $a$. Using the Pauli matrices $\sigma_{\alpha}^{(i)}$, the Hamiltonian in the
rotating frame after applying the rotating wave approximation reads
%
\begin{equation}
H=-\frac{\hbar \Delta}{2} \sum_i \sigma_z^{(i)}+\frac{\hbar \Omega}{2} \sum_i
\sigma_x^{(i)}+\frac{C_6}{a^6} \sum_{j<i}
\frac{P_{ee}^{(i)}P_{ee}^{(j)}}{(i-j)^6}
\label{eq:H}
\end{equation}
with ${P}_{ee}^{(i)}= (1+\s_z^{(i)})/2$ \cite{Robicheaux2005}. 
Incoherent processes such as
spontaneous emission are much slower than typical experimental timescales \cite{Saffman2010} and can be ignored.
In addition, it is important to note that 1D lattices suitable for Rydberg excitations can be
realized using deep optical lattices \cite{Bloch2008}, optical \cite{Urban2009,Gaetan2009} and magnetic \cite{Whitlock2009}
microtraps, or microfabricated arrays of thermal
vapor cells \cite{Kubler2010}.  We do not require to have one single
atom per lattice site; instead it is possible to work with collective
degrees of freedom where a single Rydberg excitation is shared among
all atoms of one lattice site. In the following we focus on the situation with a single atom
per lattice site.

The derivation of the phase diagram from the microscopic Hamiltonian
follows in three steps: (i) we start in the classical regime for
$\Omega=0$, and derive the devil's staircase structure of the
commensurate crystal. (ii) We study the influence of the driven
dynamics with $\Omega\neq0$ within perturbation theory and derive the
effective low energy Hamiltonian.  (iii) For increasing
number of defects, we study the effective Hamiltonian within
mean-field theory and establish the transition into the paramagnetic
phase.

For $\Omega = 0$ there are no quantum fluctuations, i.e., the
Hamiltonian is purely classical. The ground state for $\Delta > 0$
follows a complete devil's staircase of crystalline configurations
with different commensurate lattice spacings
\cite{Bak1982,Burnell2009}. The stability of each crystalline phase
with a rational filling factor $f=p/q$ is determined by the vanishing
of the single particle energy gap, i.e., when it is energetically
favorable to insert or remove one Rydberg excitation. For each $f$ we
can calculate the detuning $\Delta_0$ in the center of the lobe and
its width $\Delta_w$, which leads in the strongly interacting regime
with $f\ll 1$ to
\begin{equation}
  \Delta_0 = \frac{7\zeta(6)C_6}{a^6}\left(\frac{p}{q}\right)^6,\hspace{3em}\Delta_w = \frac{42\zeta(7)C_6}{q^{7}a^6}.
\end{equation}
Note, that the lobes with with $p>1$ are extremely small, and we will
restrict the analysis in the following on the stability of the lobes
with $p=1$, i.e., $f=1/q$.

In the following, $x_{i}$ denotes the position of the $i^{\rm th}$
Rydberg excitation in the lattice. Then, the commensurate solid
satisfies $x_{i+1}-x_{i}= q$.  Within this notation, the lowest energy
excitations are characterized that the distance between two
neighboring Rydberg excitations is reduced (increased) by a lattice
site, i.e., $x_{i+1}\!-\!x_{i} =q \mp 1$; the two types of excitations
are denoted as particle (hole) excitations, respectively. Note, that
these elementary excitations describe fractional spin excitations
\cite{Bak1982}, as the addition (removing) of a Rydberg excitation
gives rise to $q=1/f$ elementary excitations.  Their excitation
energies in the classical regime are $E_{p,h} = f \hbar \left(
\Delta_{w}/2 \!\mp\!\delta \Delta \right)$ for particles and holes,
respectively. Here, $\delta \Delta = \Delta
-\Delta_{0}$ is the deviation of the detuning from the center of the lobe. 
In addition, two excitations of equal type have a repulsive
interaction, while two excitations of different type are attractive.

\begin{figure}[tb]
  \includegraphics[width=0.9\linewidth]{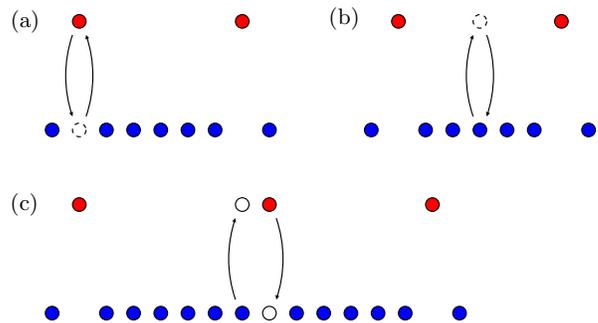}
  \caption{Processes in second order perturbation theory. (a) Virtual
    annihilation of a Rydberg excitation. (b) Virtual creation of a
    Rydberg excitation. (c) Hopping of a crystal defect.}
  \label{fig:perturb}
\end{figure}

Next, we focus on the regime with a finite drive $\Omega \neq0$, where the system acquires
a highly non-trivial dynamics. Within this regime, the melting of the commensurate crystals appears via 
the nucleation and subsequent condensation of defects. 
For a large
detuning $\Delta \gg \Omega$, the system has a well-defined energy gap
and we can derive an effective low-energy Hamiltonian for the defects
within perturbation theory in $\Omega/\Delta$. 
We first restrict
our treatment to three classes of states: (i) the crystalline ground
state $\ket{c}$, (ii) the states $\ket{p_i}$ with a particle-like
defect between the Rydberg excitation $x_i$ and $x_{i+1}$, i.e., $
x_{i+1}\!-\!x_{i}= q-1$, and (iii) the analogous states $\ket{h_i}$
for hole-like defects.  In second order perturbation theory, the
effective Hamiltonian contains diagonal terms, shown in
Fig.~\ref{fig:perturb}(a), and Fig.~\ref{fig:perturb}(b), while the
particle (hole) defects also acquire an off-diagonal term corresponding
to a hopping of the defect, see Fig.~\ref{fig:perturb}(c).
Adding a constant energy such that $E_c = 0$ the diagonal terms take
the form
\begin{equation*}
\bra{\alpha}\heff\ket{\alpha} = E_\alpha + \sum \limits_{x_{i}} \frac{\hbar^2 \Omega^2}{E_\alpha -E_i^{(a)}} + 
\sum\limits_{j\notin  \{x_{i}\}}\frac{\hbar^2\Omega^2}{E_\alpha -E_j^{(b)}}.
\end{equation*}
Here,  $E_j^{(a)}$ and $E_j^{(b)}$ correspond to the energies of the
virtual levels depicted in Fig.~\ref{fig:perturb}(a-b), which take the form
\begin{eqnarray}
E_{i}^{(a)}-E_{\alpha} & = &- \frac{C_{6}}{a^6} \sum_{x_{j}\neq x_{i}} \frac{1}{ |x_{i}-x_{j}|^6}+\hbar \Delta,\label{virtualenergies}\\  
E_{j}^{(b)} -E_{\alpha }&= & \frac{C_{6}}{a^6} \sum_{x_{i}} \frac{1}{ |x_{i}-x_j|^6}-\hbar \Delta.  \nonumber
\end{eqnarray}
Consequently, the diagonal terms provide an additional shift in the
excitation energy for the particles and holes, which in the limit
$f\ll 1$ reduces to ($\alpha = p,h$)
\begin{equation}
  \Delta E_{\alpha}  =  \bra{\alpha_{i}}\heff\ket{\alpha_{i}}- \bra{c}\heff\ket{c}-E_\alpha = I_{\alpha}\frac{\hbar \Omega^2}{\Delta},
\end{equation}
with the dimensionless quantities $I_{p}\approx -I_h \approx 0.090$.
The off-diagonal defect hopping for holes reduces to 
\begin{equation}
  J_{h}= \bra{h_i}\heff\ket{h_{i\pm1}}  = 
   \frac{\hbar^2\Omega^2}{E_h-E_i^{(c_{1})}} + \frac{\hbar^2\Omega^2}{E_h-E_i^{(c_2)}},
\end{equation}
where the first process annihilates a Rydberg excitation at position $x_{i}$ with the subsequent creation of the Rydberg excitation at position
$x_{i} \pm 1$, while the second term reverses the order of creation and annihilation.
Finally, the corresponding term $J_{p}$ for the hopping of particle defects
is obtained similarly, and the 
evaluation of the second order processes
provides $J_{p} = K_{p} \hbar \Omega^2/\Delta$ and  $J_{h} = K_{h} \hbar \Omega^2/\Delta$ 
with $K_{p} \approx K_{h} \approx -7/5$.
We can now derive the shape of the lobes for the commensurate solid, see Fig.~\ref{fig:phase}: the nucleation of particle like defects destroying the commensurate 
solid takes place at the vanishing of the excitation gap, i.e., $E_{p} + \Delta E_{p}- 2 J_{p}=0$ and determines the upper 
boundary of the lobes, while the nucleation of hole like defects becomes preferable for $E_{h} + \Delta E_{h}- 2 J_{h}=0$, 
which determines the lower boundary for the lobes. 

The new phase is characterized by a finite density of defects. These defects obey a 
strong on-site repulsion, and consequently, close to the phase transition line the density of 
defects is very low. This allows us to describe the qualitative behavior of the quantum phase transition
within an effective spin-$1$ theory for the defects
\begin{equation}
    \heff =  - J \sum_{i} \left[ S_{i}^{+} S_{i+1}^{-}+{\rm h.c.}\right]  + U  \left(S_{i}^{z}\right)^2  - \mu S_{i}^{z}
\end{equation}
with the three states $|m\rangle_{i}$ describing the presence of a particle ($m=1$) or hole ($m=-1$) 
defect at site $i$. The $XY$ interaction with $J= J_{p}\approx J_{h}$ includes the hopping of the defects and
the possibility to create particle-hole pairs, while the uniaxial anisotropy $U=f \hbar \Delta_{w}/2$ accounts for the cost in energy to 
create defects, and the chemical potential $\mu=\hbar \delta \Delta$ describes the variation in detuning.
Note, that this model neglects additional weak particle-hole symmetry breaking terms, longer range interactions, 
higher orders in $\Omega/\Delta$ perturbation theory, and the possibility to nucleate several defects at the same site. 
However, these additional terms will only change the quantitative behavior of the commensurate lobes.

\begin{figure}
  \includegraphics[width=0.9\linewidth]{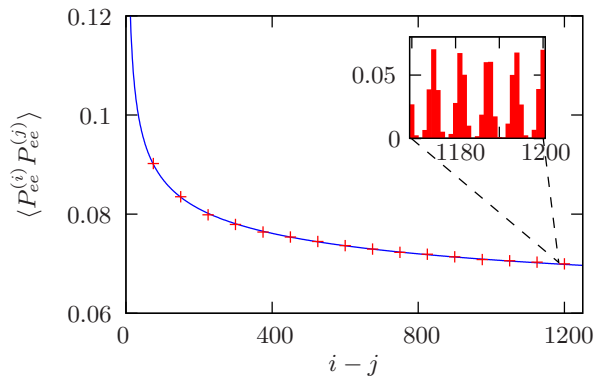}
    \caption{Decay of the spin correlations from Monte-Carlo
      simulations (crosses) according to $c/\sqrt{\ln j/b'}$ (solid
      line) as predicted by Luttinger liquid theory ($q=6$, $n =
      0.25$). The crossover to the algebraic decay takes place at much larger distances $|i-j|\sim 10^{9}$. The inset shows the oscillations of the correlation function.}
  \label{fig3}
\end{figure}

The effective one-dimensional spin-$1$ model gives naturally rise to a 
transition from the 
commensurate solid with an excitation gap to a phase with algebraic correlations 
$\langle S_{i}^{z} S_{j}^{z}\rangle - \langle S_i^z\rangle^2 \sim 1/|i-j|^{2 K}$ and linear excitation spectrum.
Varying the detuning $\delta \Delta$ results in a commensurate-incommensurate transition with defects behaving as free fermions, i.e., $K=1$, 
while the transition at the tip of the lobe is described by 
the Kosterlitz-Thouless universality class  with $K=2$ \cite{Chen2003}.
However, the physical quantity describing
the properties of the Rydberg atoms  is given by the spin-spin correlation  
$\langle P_{ee}^{(i)}P_{ee}^{(j)}\rangle$ rather than the
defect correlations $\langle S_{i}^{z} S_{j}^{z}\rangle$. 
 Consequently, we have to provide a mapping allowing to calculate the physical quantity from the effective model. The defect density $S_i^z$ can be expressed in terms of the position of the Rydberg atoms
as $ S_{i}^{z} = x_{i+1} - x_i - q$. Consequently, the total defect number 
$N_k = \sum_{j=0}^{i-1}S_{j}^{z} = x_{k}\!-\!x_{0} \!-\! k q$  between the Rydberg excitation at $x_{0}$
and $x_{k}$ defines the position of the  $k^{\rm th}$ Rydberg excitation.
As the system is translational invariant, we can assume without loss  
of generality  $x_{0}=0$. Then, the correlations between the Rydberg atoms take the form
\begin{equation}
\langle P_{ee}^{(0)}  P_{ee}^{(j)} \rangle= \frac{1}{q+n} \sum_k  \langle \delta_{j,k q +N_k}\rangle =  \sum_k  \frac{ P_{k}(j-k q)}{q+n}  \label{correlationfunction}
\end{equation}
with $n = \langle n_{i } \rangle$ the  defect density in the effective Hamiltonian.
Here,  $P_{k}(m)$ denotes the probability distribution of $N_{k}$ to find $m$ defects.
In the regime where the system is described by free fermions, the distribution
function can be determined efficiently numerically at short distances using Monte-Carlo
simulations with correlated random numbers \cite{Qaqish2003}, see Fig.~\ref{fig3}. On the other hand,
the long distance behavior can be derived within Luttinger liquid theory \cite{Haldane1981}, predicting a 
Gaussian distribution
%
%
with a mean value $n k$ and a variance $\kappa^2 = \langle (N_{k}- n k)^2 \rangle = K  \log( k /b)/\pi^2$; 
here $b\sim \pi n$
denotes a short distance cutoff.  As shown in Fig.~\ref{fig3}, this Gaussian distribution even captures the 
qualitative behavior on short distances. Consequently, we find that the Rydberg-Rydberg correlations oscillate with period $n+q$ and decay on short distances
as $1/\sqrt{\log[j/(n+q)/b]}$, while a transition to an algebraic decay occurs at large distances, i.e.,
\begin{displaymath}
    \frac{\langle P_{ee}^{(0)} P^{(j)}_{ee}\rangle-\langle P^{(0)}_{ee} \rangle^2 }{\langle P^{(0)}_{ee} \rangle^2} \sim  \cos\left( \frac{2 \pi j}{n+q}\right) \left[\frac{(n+q) b}{j}\right]^{\frac{2K}{(n+q)^2}} 
\end{displaymath}
with $\langle P^{(0)}_{ee} \rangle=1/(n+q)$.  
Therefore, we find that the transition  from the commensurate solid takes place into a novel phase exhibiting quasi long-range order
for the solid correlation function. As the density is in general incommensurate with the underlying lattice structure we denote this novel phase
as a floating solid.
It is important to note, that the crossover from the short distance regime to the algebraic decay takes place once 
the different terms in Eq.~(\ref{correlationfunction}) start to overlap, i.e., $j \sim \exp[(n+q)^2/2]$. 
While this scale is much larger than the typical
  experimental scale achievable, the short distance $1/\sqrt{\log
    j/b'}$ decay is a unique characteristic for this phase.

In the regime of large Rabi frequencies with $\Omega \gg C_6/a^6\gg
\Delta$, the spin Hamiltonian (\ref{eq:H}) reduces to the Ising model
in a transverse and a longitudinal field. Then, the
  ground state is described by a gapped paramagnetic phase with
  algebraic decay of the correlations as $1/r^6$ due to the long range
  interaction \cite{Schuch2006}. Consequently, a
  second phase transition takes place from the gapless floating solid
  to the paramagnet. 

This second transition can be seen as the breakdown of an effective theory describing the 
system in terms of defects, and takes place  when the number of defects $n_{i}$ 
at a site $i$ becomes of the order of the spacing between the Rydberg atoms $q$.  Note, that this criterion is
equivalent to  classical Lindemann criterion for the melting of solids \cite{Kleinert1989}. The effective Hamiltonian
at the commensurate density $\mu= 0$ describing the qualitative behavior for higher defect densities
includes multiple defects at a site, and also allows  for multiple defect hopping, i.e.,
\begin{displaymath}
\heff = U\sum\limits_{in}n_i^2 -J\sum_{i} \left[\sum\limits_{mnp}\ketbrap{n}{n+p}_i\otimes\ketbrap{m+p}{m}_{i+1}\right]
\end{displaymath}
with $|n\rangle_{i}$ describing the state with $n$ defects at site $i$ ($n<0$ for holes and $n>0$ for particles).
Using mean field theory, we obtain for the fluctuations of the defect 
number $\langle n_{i}^2 \rangle \sim  J^2/U^2$,
and comparing this value with the mean spacing between the Rydberg atoms  $q^2 \sim J^2/U^2$
provides the scaling $\Delta\sim \Omega^{12/13}$ for  the transition line from the floating solid towards the
paramagnet. Note, the influence of the underlying lattice drops out
for this second transition, and the scaling agrees with the value
previously derived using a universal scaling function \cite{Low2009}.
The complete phase diagram exhibiting three different phases is
sketched Fig.~\ref{fig:phase}. Note that this phase diagram is also in
good agreement with recent numerical results obtained for the spin
Hamiltonian (\ref{eq:H}) at high fillings \cite{Schachenmayer2010}.

Experimentally, the commensurate solids are the most
  challenging to realize. They can be observed if the critical Rabi
  frequency $\Omega_c$ at the tip of the lobe is much larger than the
  intrinsic decoherence rates, e.g., by radiative decay. For $^{87}\mathrm{Rb}$ atoms with a
  principle quantum number $n = 80$ and a lattice spacing $a =
  266\,\mathrm{nm}$, we find for the $q = 32$ phase $\Omega_c =
  2\pi\times 1.5\,\mathrm{MHz}$. Note that isolated radiative decays will lead to an overall reduction in Rydberg atom density, but correlations will be largely unaffected.

Finally,  the presence of a  floating solid with incommensurate fillings between the commensurate solid lobes and the
paramagnetic phase is highly remarkable as the underlying lattice structure is not required to stabilize this phase.
Consequently, we expect that this floating solid with quasi long-range order also survives a different arrangement of the atoms; especially we expect that this phase is also present in the 'frozen' Rydberg regime \cite{Anderson1998}, where the atoms are distributed randomly.

Discussions with A. Daley and J. Schachenmayer are acknowledged. The
work was supported by the Deutsche Forschungsgemeinschaft (DFG) within
SFB/TRR 21.


\begin{thebibliography}{10}

\bibitem{Kleinert1989}
H.~Kleinert,
\newblock {\em Stresses and Defects: Differential Geometry, Crystal Melting}
  Vol.~II of {\em Gauge fields in condensed matter} (World Scientific,
  Singapore, 1989).

\bibitem{Wigner1934}
E.~Wigner,
\newblock Phys. Rev. {\bf 46}, 1002 (1934);
B.~Tanatar and D.~M. Ceperley,
\newblock Phys. Rev. B {\bf 39}, 5005 (1989).

\bibitem{Buchler2007}
H.~P. B\"uchler et~al.,
\newblock Phys. Rev. Lett. {\bf 98}, 060404 (2007);
G.~E. Astrakharchik et~al.,
\newblock {\em ibid.} {\bf 98}, 060405 (2007).


\bibitem{Weimer2008a}
H.~Weimer et~al.,
\newblock Phys. Rev. Lett. {\bf 101}, 250601 (2008).

\bibitem{Pohl2010}
T.~Pohl, E.~Demler, and M.~D. Lukin,
\newblock Phys. Rev. Lett. {\bf 104}, 043002 (2010).

\bibitem{Schachenmayer2010}
J.~{Schachenmayer} et~al.,
\newblock New. J. Phys. {\bf 12}, 103044 (2010).

\bibitem{Saffman2010}
M.~Saffman, T.~G. Walker, and K.~M{\o}lmer,
\newblock Rev. Mod. Phys. {\bf 82}, 2313 (2010).

\bibitem{Weimer2010}
H.~{Weimer} et~al.,
\newblock Nature Phys. {\bf 6}, 382 (2010).

\bibitem{Heidemann2007}
R.~Heidemann et~al.,
\newblock Phys. Rev. Lett. {\bf 99}, 163601 (2007).

\bibitem{Urban2009}
E.~Urban et~al.,
\newblock Nature Phys. {\bf 5}, 110 (2009)

\bibitem{Gaetan2009}
A.~Ga\"etan et~al.,
\newblock Nature Phys. {\bf 5}, 115 (2009).

\bibitem{Younge2009}
K.~C. Younge et~al.,
\newblock Phys. Rev. A {\bf 79}, 043420 (2009).

\bibitem{Pritchard2009}
J.~D. Pritchard et~al.,
\newblock Phys. Rev. Lett. {\bf 105}, 193603 (2010).

\bibitem{Schempp2010}
H.~Schempp et~al.,
\newblock Phys. Rev. Lett. {\bf 104}, 173602 (2010).

\bibitem{Olmos2009}
B.~Olmos, R.~Gonz\'alez-F\'erez, and I.~Lesanovsky,
\newblock Phys. Rev. A {\bf 79}, 043419 (2009).

\bibitem{Low2009}
R.~L\"ow et~al.,
\newblock Phys Rev. A {\bf 80}, 033422 (2009).

\bibitem{Robicheaux2005}
F.~{Robicheaux} and J.~V. {Hern{\'a}ndez},
\newblock Phys. Rev. A {\bf 72}, 063403 (2005).

\bibitem{Bloch2008}
I.~Bloch, J.~Dalibard, and W.~Zwerger,
\newblock Rev. Mod. Phys. {\bf 80}, 885 (2008).

\bibitem{Whitlock2009}
S.~Whitlock et~al.,
\newblock New J. Phys. {\bf 11}, 023021 (2009).

\bibitem{Kubler2010}
H.~K\"ubler et~al.,
\newblock Nature Photon. {\bf 4}, 112 (2010).

\bibitem{Bak1982}
P.~Bak and R.~Bruinsma,
\newblock Phys. Rev. Lett. {\bf 49}, 249 (1982).

\bibitem{Burnell2009}
F.~J. Burnell et~al.,
\newblock Phys. Rev. B {\bf 80}, 174519 (2009).

\bibitem{Chen2003}
W.~Chen, K.~Hida, and B.~C. Sanctuary,
\newblock Phys. Rev. B {\bf 67}, 104401 (2003).

\bibitem{Qaqish2003}
B.~F. Qaqish,
\newblock Biometrika {\bf 90}, 455 (2003).

\bibitem{Haldane1981}
F.~D.~M. Haldane,
\newblock Phys. Rev. Lett. {\bf 47}, 1840 (1981).

\bibitem{Schuch2006}
N.~Schuch, J.~Cirac, and M.~Wolf,
\newblock Comm. Math. Phys. {\bf 267}, 65 (2006).

\bibitem{Anderson1998}
W.~R. Anderson, J.~R. Veale, and T.~F. Gallagher,
\newblock Phys. Rev. Lett. {\bf 80}, 249 (1998).

\end{thebibliography}


\end{document}